\documentstyle[12pt]{article}

\def\be{\begin{equation}}
\def\en{\end{equation}}
\def\bq{\begin{eqnarray}}
\def\eq{\end{eqnarray}}
\begin{document}
\begin{center}
{\large \bf Comment on "Duality of $x$ and $\psi$ in 
Quantum Mechanics" by A.E. Faraggi and M. Matone}\\[1.5cm]
\end{center}
\centerline{{\large \bf J.L. Lucio$^{(1)}$ and M. Ruiz--Altaba$^{(2)}$}}
\vspace{.8cm}
\centerline {{\it $^{(1)}$ Instituto de F\'\i sica, 
Universidad de Guanajuato, Le\'on, Gto., M\'exico}}
\centerline{{\it $^{(2)}$ Instituto de F\'\i sica, 
UNAM, A.P. 20-364, 01000 M\'exico, D.F.}}
\vspace{.8cm}
Recently, Faraggi and Matone introduced [1] an equation ``dual" to the 
time--independent Schr\"odinger equation 
\be 
{ - {\hbar^2 \over 2m}{ d^2 \psi \over dx^2} = \left[ E- V(x) \right] \psi ,}
\en
namely
\be { {\hbar^2 \over 2m} {\partial^2 x \over \partial \psi ^2} =
 \left[E-V(x) \right] \psi \left( \partial x \over \partial \psi \right)^3 ,}
\en
which they derived [1] in a rather roundabout way with the help of a 
``prepotential''  $\cal F(\psi)$. They went on to speculate on a possible dual
quantization of space based on this formula, employing also this formalism to 
study duality transformations in the pilot--wave version of quantum mechanics 
[2].

We would like to point out that, regardless of the conceptual interest and 
difficulties of interpreting space in terms of a wave--function instead of the
opposite, equation (2) can be derived very easily from the trivial observation
that  
\be
\frac{d\psi}{dx} = \left[\frac{dx}{d\psi}\right]^{-1}  ,
\en
and thus 
\be
{ {d^2 x \over d\psi^2 } = - {d^2 \psi \over dx^2 } \left( dx \over d\psi 
\right)^3 .}
\en

Plugging (4) into (1) gives (2) directly.  A more interesting truly dual 
version of (1) should involve a dual potential $\tilde V (\psi)$; we have no 
idea how to derive such an equation, nor what its interpretation would be.

{\bf Acknowledgements}

Work partially supported by CONACyT under contract 3979PE-9608 and Catedra 
Patrimonial de Excelencia Nivel II.

\vspace{.5cm}
{\bf Bibliography}

[1] {A.E. Faraggi and M. Matone, {\sl Phys. Rev. Lett.} 
{\bf    78}  (1997)    163; {\tt hep-th/9606063}.}

[2] {A.E. Faraggi and M. Matone, {\tt hep-th/9705108}.}. 

\end{document}